\documentclass[journal]{IEEEtran}

\usepackage{amsmath,graphicx}
\usepackage{algorithm}
\usepackage{algorithmic}

\usepackage{amsfonts}
\usepackage{amssymb}
\usepackage{bbm}
\usepackage{accents}
\usepackage{comment}

\begin{document}
\title{MIMO Channel Prediction via Deep Learning-based Conformal Bayes Filter}

\author{Dongwon Kim,~\IEEEmembership{Student Member,~IEEE,}
        Jinu Gong,
        and~Joonhyuk~Kang,~\IEEEmembership{Member,~IEEE}% <-this % stops a space
\thanks{Dongwon Kim and Joonhyuk Kang are with the School of Electrical Engineering, Korea Advanced Institute of Science and Technology (KAIST), Daejeon 34141, South Korea (e-mail: davinci5694@kaist.ac.kr; jhkang@ee.kaist.ac.kr).\\
Jinu Gong is with the Department of Applied Artificial Intelligence, Hansung University, Seoul, Korea (e-mail: jinugong@hansung.kr)}

\thanks{This work was partly supported by the Institute of Information \& Communications Technology Planning \& Evaluation (IITP)-ITRC (Information Technology Research Center) grant funded by the Korea government (MSIT) (IITP-2025-RS-2020-II201787, contribution rate: 50\%); in part by the Institute of Information \& Communications Technology Planning \& Evaluation (IITP) under 6G$\cdot$Cloud Research and Education Open Hub grant funded by the Korea government (MSIT) (IITP-2025-RS-2024-00428780, contribution rate: 25\%). Also, this research was financially
supported by Hansung University for Jinu Gong (contribution rate: 25\%).  }}

% The paper headers
\markboth{IEEE WIRELESS COMMUNICATIONS LETTERS,~Vol.~XX, No.~XX, XXX~2026}%
{Shell \MakeLowercase{\textit{et al.}}: Bare Demo of IEEEtran.cls for IEEE Journals}

\maketitle

% As a general rule, do not put math, special symbols or citations
% in the abstract or keywords.
\begin{abstract}
Channel prediction has emerged as an effective solution for acquiring accurate channel state information (CSI) in the presence of channel aging. Existing methods have inherent limitations, with conventional Kalman filter (KF)-based approach being vulnerable to model mismatch and deep learning (DL)-based approaches producing overconfident predictions. To address these issues, we propose a DL-based conformal Bayes filter (DCBF) that integrates DL-based prediction, conformal quantile regression (CQR), and Bayesian filtering. The proposed framework enables principled fusion of calibrated priors and observations, yielding reliable channel predictions with the calibrated uncertainty. Simulation results demonstrate that DCBF significantly improves DL-based prediction and outperforms the KF–based method.
\end{abstract}

% Note that keywords are not normally used for peerreview papers.
\begin{IEEEkeywords}
MIMO channel prediction, Conformal prediction, uncertainty quantification, Bayesian filtering
\end{IEEEkeywords}

\IEEEpeerreviewmaketitle

\section{Introduction}
\IEEEPARstart{M}{ultiple}-input multiple-output (MIMO) system is a cornerstone of modern wireless communication networks, as it significantly enhances spectral efficiency through spatial multiplexing and diversity gains \cite{6798744, 10054381}. 
Realization of these benefits calls for accurate channel state information (CSI), which enables efficient beamforming and precoding to maximize the system performance. However, acquiring timely CSI remains a critical challenge. The increasing number of antennas imposes substantial pilot overhead, while user mobility causes the channel aging in which the estimated channel rapidly becomes outdated \cite{channel2015aging}. To overcome these issues, channel prediction has emerged as a promising solution. Since wireless channels exhibit temporal correlation due to the quasi-stationarity of the scattering environment \cite{5947055}, future channel states can be predicted from past observations.

The conventional channel prediction typically relies on the statistical approaches that approximate fading channels as an autoregressive (AR) process and apply Kalman filtering (KF) under linear model assumption, which often does not hold in practice \cite{min2007mimo, kashyap2017performance}. Moreover, the computational complexity of KF scales cubically with the number of antennas, which incurs significant computational overhead that restricts its use for low-latency prediction in MIMO systems.  

With the substantial advancement of deep learning (DL), DL-driven channel prediction methods have been actively explored for MIMO systems \cite{jiang2020deep, ko2025machine}. Leveraging the strong representation capability of the deep neural network (DNN), DL models can effectively learn complex channel characteristics from large-scale datasets. For instance, convolutional neural networks (CNNs) have been employed to exploit the spectral correlations for channel reconstruction \cite{8979256}, while long short-term memory (LSTM) networks capture temporal dependencies across successive channel realizations \cite{helmy2023lstm}. More recently, transformer-based architectures have been introduced for channel prediction \cite{jiang2022accurate}, leveraging a self-attention mechanism to model long-range dependencies and further improve accuracy. Despite these advances, most DL-based channel prediction frameworks provide only point estimates, which often lead to overconfident and unreliable predictions. 

In contrast to prior work, our approach integrates reliable uncertainty quantification into DL-based channel prediction, enabling statistically validated predictions and mitigating the overconfidence inherent to conventional DL-based predictors. To achieve calibrated uncertainty quantification, we use a conformal quantile regression (CQR), which adjusts the raw quantile predicts of the predictor using calibration data so that adjusted quantiles satisfy guaranteed coverage properties. Several CP works have been recently applied to wireless network systems such as resource allocation \cite{10964108}, prediction \cite{zecchin2025generalization}, and beamforming selection \cite{deng2025scan}.   

In this work, we develop a deep conformal Bayes filter (DCBF) built upon the DL-based channel predictor. The proposed framework consists of three steps. First, we use the DNN model as a quantile predictor to estimate the quantiles of future channels. Second, we adopt CQR to calibrate quantiles, leading to a well-calibrated predictive distribution. Finally, Bayesian filtering incorporates the calibrated predictive distribution with noisy observations to recursively refine channel predictions. Simulation results on various channel models demonstrate that the proposed DCBF achieves superior performance over KF-based and DL-based channel predictors.

The remainder of the paper is organized as follows. Section \ref{sec:system_model} formulates the channel prediction problem. Section \ref{sec:methodology} introduces the DCBF framework. Section \ref{sec:simulation_results} evaluates the performance of the proposed method with other baselines. Finally, Section \ref{sec:conclusion} concludes this paper.
\begin{figure*}[!t]
     \centering
    \includegraphics[width=\textwidth]{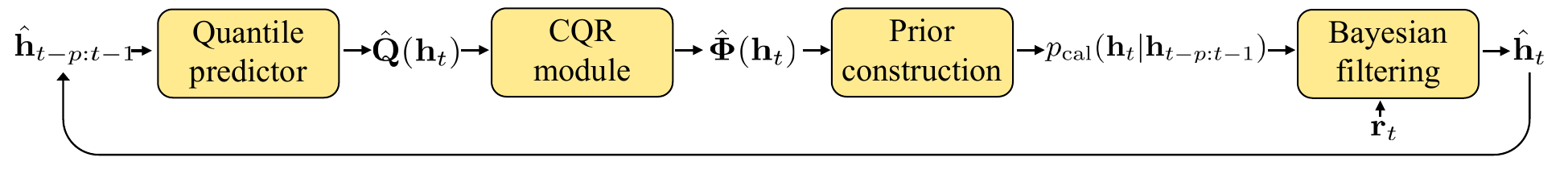}
    \caption{CBF framework: (i) prediction step through DL-based channel quantile predictor (ii) calibration via CQR (iii) prior distribution construction with calibrated quantiles (iv) integration of prediction and measurement information through Bayesian filtering.}
    \label{fig:Illustration_of_CBF_framework}
\end{figure*}
\section{System Model}
\label{sec:system_model}

We consider a single-cell MIMO system where the base station (BS) is equipped with $M$ antennas and each of the $K$ user equipments (UEs) is equipped with  $N$ antennas. Orthogonal frequency division multiple access 
(OFDMA) is assumed, which cancels inter-user interference. Consequently, channel prediction is performed by focusing on the single UE without loss of generality. The received signal at the BS from the $k$-th UE at time slot $t$ is given by
\begin{align}
\label{eq:system_model_received_signal}
    \mathbf{r}_{t,k} &= \sqrt{\rho}\mathbf{H}_{t,k}\mathbf{s}_{t,k}+ \mathbf{w}_{t,k},
\end{align}
where $\mathbf{H}_{t,k} \in \mathbb{C}^{M\times N}$ is the channel matrix between BS and the $k$-th UE, $\rho$ is the transmit power, $\mathbf{s}_{t,k} \in \mathbb{R}^{N\times1}$ is the transmitted signal and $\mathbf{w}_{t,k} \sim \mathcal{CN}(\mathbf{0}_M, \sigma^2_w\mathbf{I}_M)$ is complex additive white Gaussian noise (AWGN). Note that the user index $k$ will be omitted for simplicity. 

The channel prediction problem is formulated based on the $p$ previous received signals $\{\mathbf{r}_{t-p+1},\ldots,\mathbf{r}_t\}$ for $p\geq1$, which can be formulated as
\begin{align}
    &\min\, ||\mathbf{H}_{t+1} - \hat{\mathbf{H}}_{t+1}||^2,\\ 
    &\text{subject to\,} \hat{\mathbf{H}}_{t+1} = f(\mathbf{r}_{t-p+1},\ldots,\mathbf{r}_t),
\end{align}
where $\hat{\mathbf{H}}_{t+1}$ denotes the predicted channel at $(t+1)$-th time slot, and $f(\cdot)$ represents the arbitrary channel predictor function. Furthermore, we vectorize the channel matrix $\mathbf{H}_t$
as $\mathbf{h}_t \in \mathbb{R}^{L\times1}$, where $L=2MN$.

\section{Conformal Bayes Filter-based Channel Prediction}
\label{sec:methodology}
In this section, we introduce the DL-based channel prediction framework that integrates a CQR module with Bayesian filtering for sequential refinement, as outlined in Fig. \ref{fig:Illustration_of_CBF_framework}. The framework consists of three main components: (i) DL-based channel quantile predictor that estimates quantiles of the future channel using DNN, (ii) CQR module that adjusts the predicted quantiles to ensure coverage guarantee, and (iii) Bayesian filtering that incorporates calibrated quantiles with the measurement information from the received signal.  
\subsection{Deep Quantile Channel Predictor}
The entire dataset is partitioned into four disjoint subsets, denoted by $\mathcal{I}_1, \mathcal{I}_2, \mathcal{I}_3, \mathcal{I}_4$, corresponding to the training set, validation set, calibration set, and test set, respectively. Using the training set $\mathcal{I}_1$, the DNN channel predictor $\hat{f}_\theta(\cdot)$, parameterized by $\theta$, is designed to predict multiple quantiles of the future channel distribution $\hat{\mathbf{Q}}(\mathbf{h}_t)$. Instead of training a separate neural network for each quantile level, a single neural network outputs a quantile vector jointly as
\begin{align}
    \hat{\mathbf{Q}}(\mathbf{h}_t)  &= \hat{f}_{\theta}(\hat{\mathbf{h}}_{t-p:t-1})\in \mathbb{R}^{ L \times G},
    \label{eq:quantile_outputs}
\end{align}
where $\hat{\mathbf{h}}_{t-p:t-1}=[\hat{\mathbf{h}}_{t-p}^{\mathrm{T}},\ldots,\hat{\mathbf{h}}_{t-1}^{\mathrm{T}}]^{\mathrm{T}}$ denotes the concatenated vector of the estimated channel $\hat{\mathbf{h}}_t$ over the past $p$ times steps and $G$ denotes the number of quantile levels. The output $\hat{\mathbf{Q}}(\mathbf{h}_t)$ represents the predicted quantiles of the future channel $\mathbf{h}_t$, where each entry $\hat{q}_{l,\tau}(\mathbf{h}_t)$ corresponds to the $\tau$-th quantile of the $l$-th antenna channel link. The parameters of DNN channel predictor $\theta$ are optimized by minimizing the aggregated quantile regression loss. For a quantile level $\tau$ and a $l$-th antenna, a pinball loss is defined as
\begin{align}
\label{eq:pinball_loss}
&\rho_{\tau}(\mathbf{h}_t[l], \hat{q}_{l,\tau}(\mathbf{h}_t)) \nonumber\\
&= \tau (\mathbf{h}_t[l]-\hat{q}_{l,\tau}(\mathbf{h}_t))\mathbbm{1}(\hat{q}_{l,\tau}(\mathbf{h}_t)\leq \mathbf{h}_t[l]) \nonumber\\
&+ (1-\tau)(\hat{q}_{l,\tau}(\mathbf{h}_t)-\mathbf{h}_t[l])\mathbbm{1}(\hat{q}_{l,\tau}(\mathbf{h}_t)>\mathbf{h}_t[l]),
\end{align}
where $\mathbbm{1}(\cdot)$ represents the indicator function. The overall loss aggregated across all antenna links and quantile levels is given by 
\begin{align}
    L(\theta) = \frac{1}{|\mathcal{I}_1|LG}\sum_{t=1}^{|\mathcal{I}_1|}\sum_{l=1}^{L} \sum_{j=1}^{G}\rho_{\tau_j}(\mathbf{h}_t[l], \hat{q}_{l,\tau_j}(\mathbf{h}_t)),
\end{align}
where $|\mathcal{I}_1|$ is the total number of training samples. The pinball loss imposes asymmetric penalties for the overestimation and underestimation according to the quantile level $\tau$, enabling the trained predictor to approximate the desired quantile. 

\subsection{Conformal Quantile Regression}
In this subsection, we explain the application of CQR to calibrate the quantile outputs in eq. \eqref{eq:quantile_outputs}. Using the calibration dataset $\mathcal{I}_3$, the conformity scores are evaluated to adjust the raw quantile prediction $\hat{\mathbf{Q}}(\mathbf{h}_t)$. For the each calibration set sample $t\in\mathcal{I}_3$, quantile level $\tau_j$, and each antenna link $l$, a conformity score is defined as
\begin{align}
e_{l,\tau_j}^{(t)}
=
\mathbf{h}_{t}[l]
-
\hat{q}_{l,\tau_j}(\mathbf{h}_t): t \in \mathcal{I}_3,
\end{align}
which measures the difference between the $\tau_j$-th quantile of prediction $\hat{q}_{l,\tau_j}(\mathbf{h}_t)$ and the true future channel $\mathbf{h}_t[l]$. CQR adjusts the raw quantile estimates as
\begin{align}
    \label{eq:cqr}
    &\hat{\phi}_{l,\tau_j}(\mathbf{h}_t)
=
\hat{q}_{l,\tau_j}(\mathbf{h}_t)
+
\widehat{\gamma},\,
\widehat{\gamma}
=
\operatorname{Quantile}_{\tau_j}
(
\, e_{l,\tau_j}^{(t)} 
,\, \mathcal{I}_3
),
\end{align}
where $\text{Quantile}_{\tau_j}(\cdot,\cdot)$ denotes the $\tau_j$-th empirical quantile over the calibration scores $e_{l,\tau_j}^{(t)}$. Under the independent and identically distributed (i.i.d.) sampling assumption that the samples of the calibration set and the test set are drawn from the same unknown distribution, the adjusted quantile satisfies the marginal coverage guarantee
\begin{align}
    \text{Pr}\left[\mathbf{h}_t[l] \leq \hat{\mathbf{\phi}}_{l,\tau_j}(\mathbf{h}_t) \right] \geq \tau_j,
\end{align}
as established in \cite{romano2019conformalized}. In other words, the adjusted quantile $\mathbf{\phi}_{l,\tau_j}(\mathbf{h}_t)$ achieves the target quantile level $\tau_j$, ensuring the true channel value lies below it with the probability at least $\tau_j$.

Once the raw quantiles are calibrated, the vector of calibrated quantiles $\hat{\mathbf{\Phi}}(\mathbf{h}_t)=[\mathbf{\phi}_{1,\tau_j}(\mathbf{h}_t)\ldots,\mathbf{\phi}_{L,\tau_j}(\mathbf{h}_t)]$ constructs the calibrated prior distribution $p_{\text{cal}}(\mathbf{h}_t|\mathbf{h}_{t-p:t-1})$. Specifically, the prior distribution for the future channel for $l$-th dimension is modeled as a piecewise uniform distribution over intervals between successive calibrated quantiles:
\begin{align}
\centering
&p_{\text{cal}}(\mathbf{h}_t[l] \mid \mathbf{h}_{t-p:t-1}[l])= \nonumber\\ 
&\begin{cases}
\dfrac{\tau_1\! -\! \tau_0}{\hat{\phi}_{l,\tau_1}(\mathbf{h}_t)\! -\! \hat{\phi}_{l,\tau_0}(\mathbf{h}_t)}, 
\text{if }\, \mathbf{h}_t[l]\! \in \! [\hat{\phi}_{l,\tau_0}(\mathbf{h}_t), \hat{\phi}_{l,\tau_1}(\mathbf{h}_t)] \\
\dfrac{\tau_2\! -\! \tau_1}{\hat{\phi}_{l,\tau_2}(\mathbf{h}_t)\! -\! \hat{\phi}_{l,\tau_1}(\mathbf{h}_t)}, 
\text{if }\, \mathbf{h}_t[l]\! \in \![\hat{\phi}_{l,\tau_1}(\mathbf{h}_t), \hat{\phi}_{l,\tau_2}(\mathbf{h}_t)] \\
\quad \vdots \\
\dfrac{\tau_{G+1}\! -\! \tau_{G}}{\hat{\phi}_{l,\tau_{G+1}}(\mathbf{h}_t)\! -\! \hat{\phi}_{l,\tau_{G}}(\mathbf{h}_t)}, 
\text{if }\, \mathbf{h}_t[l]\! \in \![\hat{\phi}_{l,\tau_{G}}(\mathbf{h}_t), \hat{\phi}_{l,\tau_{G+1}}(\mathbf{h}_t)]  \\
0, \text{ otherwise,}
\end{cases}
\label{eq:calibrated_prior}
\end{align}
where $\tau_{G+1}$ and $\tau_0$ correspond to $1$ and $0$. This construction assigns a probability mass of $\tau_g - \tau_{g-1}$ to each interval between successive quantiles, ensuring that the calibrated prior distribution is consistent with the target coverage across all quantile levels. For implementation simplicity, the predicted channel values $\hat{\phi}_{l,\tau_0}(\mathbf{h}_t)$ and $\hat{\phi}_{l,\tau_{G+1}}(\mathbf{h}_t)$ corresponding to the boundary quantile levels are set to the minimum and maximum channel values observed in the training set.

\subsection{Posterior mean estimation via importance sampling}
Given the received signal $\mathbf{r}_t$ at the $t$-th time slot, the objective is to estimate the posterior mean of the channel $\mathbf{h}_t$ that integrates the prior knowledge of the channel $\mathbf{h}_t$ from the calibrated deep quantile channel predictor and observation information from the received signal $\mathbf{r}_t$. The posterior mean for the $l$-th antenna link $\mathbf{\hat{h}}_t[l]$ is represented as
\begin{equation}
    \mathbb{E}_{p(\mathbf{h}_t[l] \mid \mathbf{r}_t[l])}[\mathbf{h}_t[l]].
    \label{eq:post_mean}
\end{equation}

According to Bayes’ rule, the posterior distribution is expressed as
\begin{align}
p(\mathbf{h}_t[l] \mid \mathbf{r}_t[l]) = \frac{p(\mathbf{r}_t[l] \mid \mathbf{h}_t[l]) \, p(\mathbf{h}_t[l])}{p(\mathbf{r}_t[l])} \label{eq:posterior_bayes},
\end{align}
where the prior distribution $p(\mathbf{h}_t[l])$ can be obtained by the calibrated distribution $p_{\text{cal}}(\mathbf{h}_t[l]|\mathbf{h}_{t-p:t-1}[l])$.
By substituting $p(\mathbf{h}_t[l])$ with $p_{\text{cal}}(\mathbf{h}_t[l]|\mathbf{h}_{t-p:t-1}[l])$ into eq. \eqref{eq:posterior_bayes}, the posterior mean becomes
\begin{align}
\label{eq:post_mean_integration}
\mathbb{E}_{p(\mathbf{h}_t[l] \mid \mathbf{r}_t[l])}[\mathbf{h}_t[l]] =& \frac{1}{p(\mathbf{r}_t[l])} \int \mathbf{h}_t[l] \, p(\mathbf{r}_t[l] \mid \mathbf{h}_t[l]) \,  \nonumber\\& \times p_{\text{cal}}(\mathbf{h}_t[l]|\mathbf{h}_{t-p:t-1}[l]) \, \mathrm{d}\mathbf{h}_t[l].
\end{align}
Since the integration in eq. \eqref{eq:post_mean_integration} is generally intractable, we approximate the integration using importance sampling with draws
$\{\mathbf{h}^{(i)}_t[l] \}_{i=1}^S \stackrel{\text{i.i.d.}}{\sim}p_{\text{cal}}(\mathbf{h}_t[l]|\mathbf{h}_{t-p:t-1}[l])$. Accordingly, the posterior mean is approximated by a weighted average of the samples as
\begin{align}
\label{eq:weighted_average}
\mathbb{E}_{p(\mathbf{h}_t[l] \mid \mathbf{r}_t[l])}[\mathbf{h}_t[l]] \approx \sum_{i=1}^{S}\gamma_i \mathbf{h}_t^{(i)}[l], 
\end{align}
where the normalized importance weight $\gamma_i$ is given as
\begin{align}
\label{eq:weights}
\gamma_i = \frac{p(\mathbf{r}_t[l] \mid \mathbf{h}_t^{(i)}[l])}{\sum_{j=1}^{S}p(\mathbf{r}_t[l] \mid \mathbf{h}^{(j)}_t[l])}.
\end{align}
Furthermore, the importance weight $\gamma_i$ is calculated by using a likelihood of each sample. For a sample $\mathbf{h}_t^{(i)}[l]$, the likelihood of the observed received signal is represented as
\begin{align}
\label{eq:likelihood}
p(\mathbf{r}_t[l]|\mathbf{h}_{t}[l]
) \sim \mathcal{N}(\sqrt{\rho}\mathbf{h}_t[l],\sigma^2_w/2),
\end{align}
where the received signal at each antenna link $l$ follows a Gaussian distribution with mean $\sqrt{\rho}\mathbf{h}_t[l]$ and and variance $\sigma^2_w/2$. The posterior mean represents the refined channel estimate and is subsequently used as the input to the DL-based channel predictor $\hat{f}_{\theta}(\cdot)$. Based on this input, the predictive distribution of the future channel is derived according to eq. \eqref{eq:calibrated_prior}. The future channel value is given by the mean of the predictive distribution, which can be straightforwardly calculated since the prior distribution is modeled as a piecewise uniform distribution.
\begin{algorithm}[t]
\caption{Channel Prediction via Deep Conformal Bayes Filter (DCBF)}
\begin{algorithmic}[1]
\label{alg:DCBF}
\REQUIRE Dataset $\mathcal{I}_1, \mathcal{I}_2, \mathcal{I}_3, \mathcal{I}_4$, quantile levels $\{\tau_j\}_{j=1}^G$, history length $p$, number of samples $S$

\FOR{$t=p+1,\ldots, T$}
    \STATE Quantile of prior $\hat{\mathbf{Q}}(\mathbf{h}_t)$ prediction via DL-based channel predictor $\hat{f}_{\theta}(\cdot)$
    \FOR{$l=1,\ldots,L$}
        \STATE Obtain the calibrated quantiles $\hat{\phi}_{l,\tau}(\mathbf{h}_t)$ as in eq. \eqref{eq:cqr}
         \STATE Construct the predictive prior distribution $p_{\text{cal}}(\mathbf{h}_t[l] \mid \mathbf{h}_{t-p:t-1}[l])$ as in eq. \eqref{eq:calibrated_prior}
         \STATE Obtain the posterior mean $\mathbb{E}_{p(\mathbf{h}_t[l] \mid \mathbf{r}_t[l])}[\mathbf{h}_t[l]]$ as in eq. \eqref{eq:weighted_average}
         \STATE Predict the next-slot channel $\hat{\mathbf{h}}_{t+1}$ using the DL-based channel predictor $\hat{f}_{\theta}(\cdot)$, where the input is the past channel history $\hat{\mathbf{h}}_{t-p+1:t}[l]$ 
     \ENDFOR
\ENDFOR
\ENSURE Return the collection of predicted channel $\{\hat{\mathbf{h}}_t\}_{t=p+2}^{T+1}$
\end{algorithmic}
\end{algorithm}

In summary, the posterior mean is calculated as a likelihood-weighted average of calibrated prior samples and serves as a refined channel estimate at time slot $t$. This value of posterior mean is subsequently fed into the DL-based channel predictor $\hat{f}_{\theta}(\cdot)$ for the future channel prediction, and the process is applied recursively to predict the channel as outlined in Algorithm~\ref{alg:DCBF}.

\section{Simulation Results}
\label{sec:simulation_results}
In this section, we compare the proposed method with several baselines including the outdated channel estimation as in \cite{kim2020massive}, the KF-based predictor, and the DL-based predictors implemented using a multi-layer perceptron (MLP) and a gated recurrent unit (GRU). The prediction accuracy is measured using the normalized mean square error (NMSE) metric defined as 
\begin{align}
    \text{NMSE} = \mathbb{E}\left[
\frac{ \| \hat{\mathbf{h}}_{t+1} - \mathbf{h}_{t+1} \|_2^2 }{ \| \mathbf{h}_{t+1} \|_2^2  }\right].
\end{align}
Overall hyperparameters of the channel model and DNN architecture are summarized in Table \ref{table:table1}. A $2\times2$ MIMO configuration is assumed, and both Urban Micro (UMi) and Urban Macro (UMa) channel models are considered \cite{specification3GPP}. The channel data are generated by using Sionna \cite{hoydis2023sionna}. For training the DL-based predictors, the batch size is set to 64 and  learning rate of $1e-3$. A total of $10000$ channel samples are generated, and the dataset is split according to the ratio $|\mathcal{I}_1|:|\mathcal{I}_2|:|\mathcal{I}_3|:|\mathcal{I}_4| = 7:1:1:1$. The predictor quantiles are set to $\tau_1,\ldots,\tau_G = [0.1,\ldots,0.9]$, with $G=9$.
\begin{table}[t]
    \vfill
    \centering
    \caption{Simulation parameters}
    \resizebox{1.0\columnwidth}{!}{
    \begin{tabular}{||c|c|c|c||} 
        \hline
        Parameters  & Value & Parameters & Value \\ [0.5ex] 
        \hline
        User speed & $5, 20$ km/h & Channel history length $p$ & 3 \\ 
        \hline
        Thermal noise & $-174$ dBm/Hz & Hidden layer size of MLP & 2 \\
        \hline
        Bandwidth & $10$ MHz & Hidden layer dimension of MLP & 64 \\
        \hline
        Carrier frequency & $3.59$ GHz  & Hidden layer size of GRU & 2 \\
        \hline
        Symbol interval & $2$ ms & Hidden layer dimension of GRU & 128 \\
        \hline
        Prediction interval & $2$ ms & Training epochs & 50\\
        \hline
    \end{tabular}}
    \label{table:table1}
\end{table}
To enhance the robustness under varying transmit power conditions, the DL-based predictors are trained with the input channel vectors perturbed by the random Gaussian noise. For a fair comparison, the KF-based predictor is also trained using the same number of samples to estimate the AR parameters, matching the dataset size used for training the DL-based predictors. 

\begin{figure}[t]
    \centering
    \includegraphics[width=0.5\textwidth]{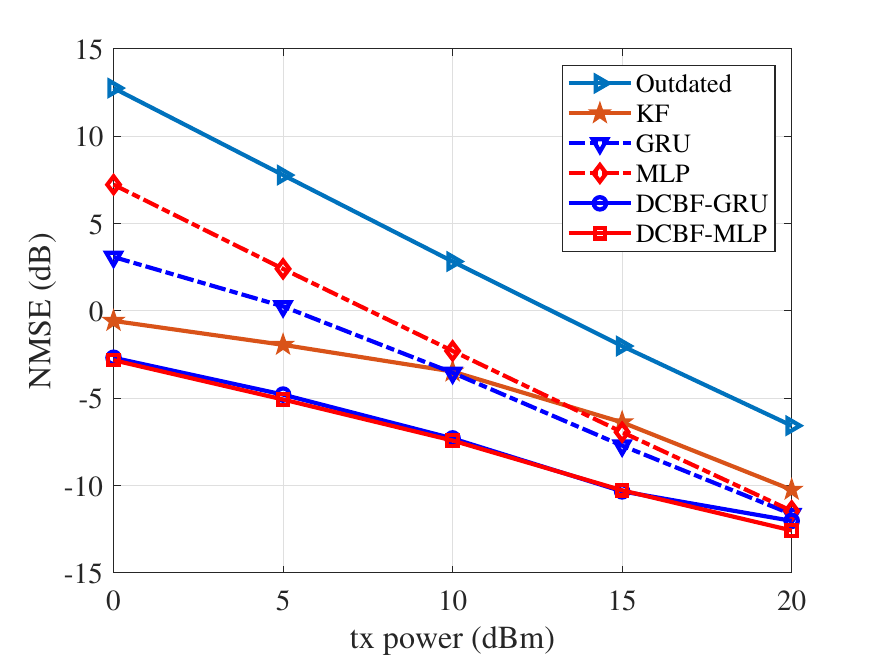} 
    \caption{NMSE of the outdated channel, the KF-based predictor, the GRU-based predictor (with and without the DCBF scheme), and the MLP-based predictor (with and without the DCBF scheme) as a function of the transmit power under the UMi channel model with a user speed of $5$ km/h.}
    \label{fig:UMi_speed_5}
\end{figure}

In Fig. \ref{fig:UMi_speed_5}, we report the NMSE as a function of transmit power where the channel model is the UMi channel model and the user speed is $5$ km/h. As the transmit power increases, the prediction error decreases for all methods. The proposed methods DCBF-MLP and DCBF-GRU outperform the KF baseline across the entire range of transmit power, achieving approximately 2–3 dB gain, and all methods converge at the transmit power of 20 dBm. Moreover, both DCBF variants improve their corresponding DL-based predictors. Note that the proposed DCBF framework yields a larger performance improvement for the MLP model. This indicates that DCBF more effectively compensates for less confident prior predictions. While the GRU model inherently outperforms the MLP, it tends to produce overconfident predictions, thereby limiting the impact of calibration and newly observed information in the filtering update.

\begin{figure}[t]
    \centering
    \includegraphics[width=0.5\textwidth]{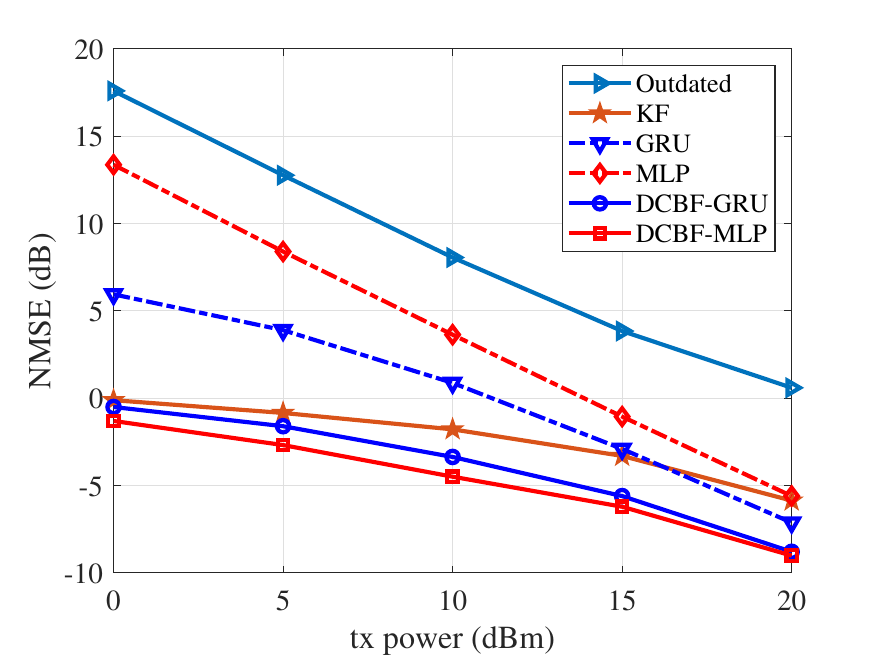} 
    \caption{NMSE of the outdated channel, the KF-based predictor, the GRU-based predictor (with and without the DCBF scheme), and the MLP-based predictor (with and without the DCBF scheme) as a function of the transmit power under the UMi channel model with a user speed of $20$ km/h.}
    \label{fig:UMi_speed_20}
\end{figure}

Fig.~\ref{fig:UMi_speed_20} shows the NMSE as a function of transmit power where the simulation environments are the UMi channel model and the user speed of $20$ km/h. In comparison with Fig.~\ref{fig:UMi_speed_5}, the overall trend remains similar to the $5$ km/h case. In particular, DCBF-MLP achieves approximately a $2.8$ dB improvement over the KF-based prediction at a transmit power of $10$ dBm.

\begin{figure}[t]
    \centering
    \includegraphics[width=0.5\textwidth]{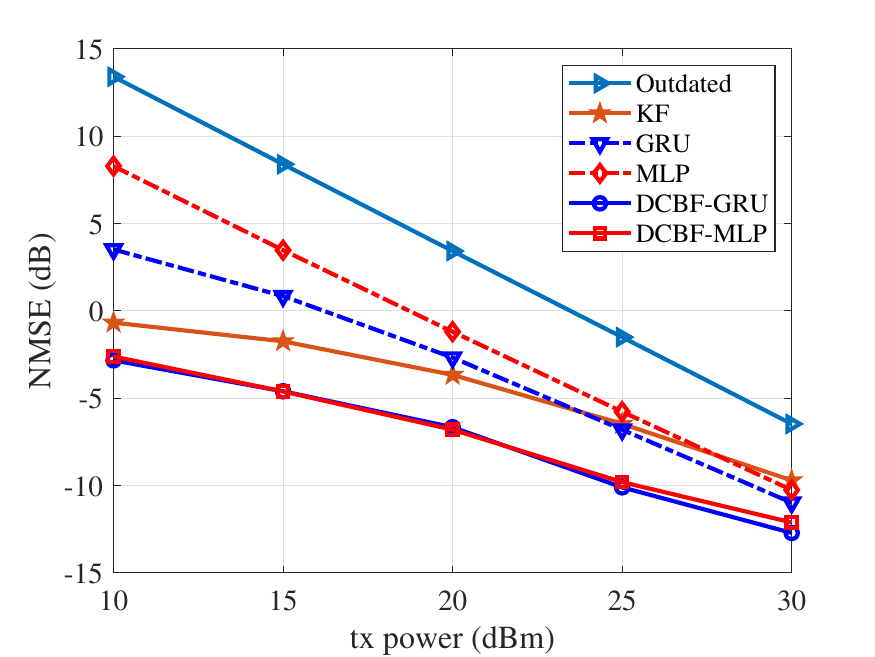} 
    \caption{NMSE of the outdated channel, the KF-based predictor, the GRU-based predictor (with and without the DCBF scheme), and the MLP-based predictor (with and without the DCBF scheme) as a function of the transmit power under the UMa channel model with a user speed of $5$ km/h.}
    \label{fig:UMa_speed_5}
\end{figure}

\begin{figure}[t]
    \centering
    \includegraphics[width=0.5\textwidth]{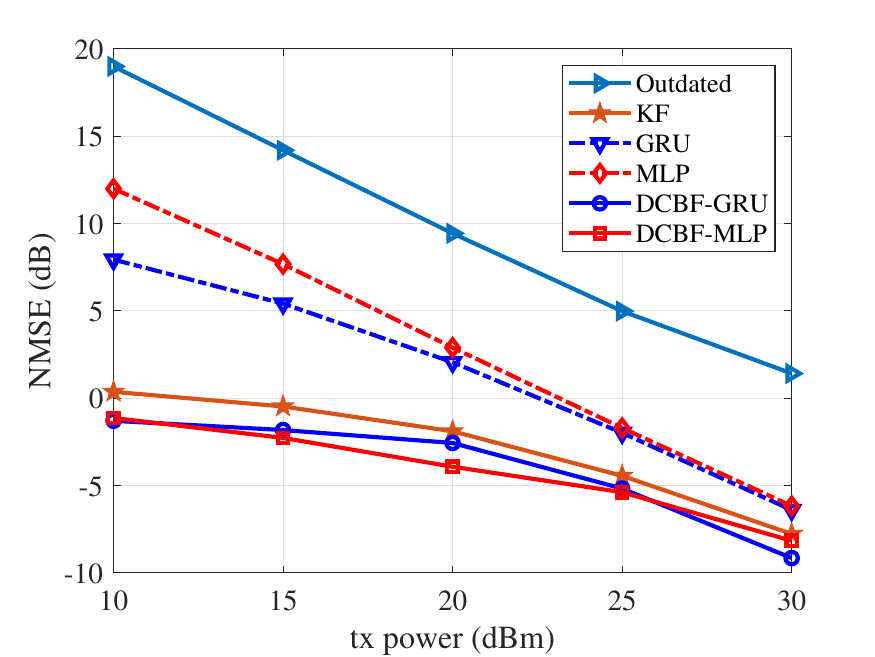} 
    \caption{NMSE of the outdated channel, the KF-based predictor, the GRU-based predictor (with and without the DCBF scheme), and the MLP-based predictor (with and without the DCBF scheme) as a function of the transmit power under the UMa channel model with a user speed of $20$ km/h.}
    \label{fig:UMa_speed_20}
\end{figure}

Fig. \ref{fig:UMa_speed_5} and Fig. \ref{fig:UMa_speed_20} compare the NMSE performance for user speeds of $5$ km/h and 20 km/h where the channel model is the UMa channel. In both channel environments, the trends of performance are similar to the UMi case. Across all transmit power levels, the proposed DCBF variants consistently outperform their non-calibrated counterparts (GRU and MLP), demonstrating the benefit of conformal calibration in constructing a more accurate prior for Bayesian filtering. These results show that DCBF remains robust across different mobility conditions and channel models, providing superior prediction accuracy compared to all baseline methods.


\section{Conclusion}
\label{sec:conclusion}
In this paper, we proposed a novel DCBF framework for MIMO channel prediction. The DCBF framework integrates the quantiles from the DL-based predictor, conformal calibration, and Bayesian filtering to produce calibrated predictive distributions and achieve principled fusion of prior predictions with noisy observations. Numerical evaluations across various 3GPP channel models and user velocities demonstrate that the proposed method consistently outperforms both KF-based predictors and existing DL-based approaches. For future work, extensions incorporating online conformal calibration to better adapt to rapidly evolving channel statistics would be of particular interest. 
\bibliographystyle{IEEEtran}
\bibliography{refs}

@inproceedings{ko2025machine,
  title={Machine {L}earning-{B}ased {C}hannel {P}rediction with {R}educed {T}raining {O}verhead for {M}assive {MIMO}-{OFDM} {S}ystems},
  author={Ko, Beomsoo and Kim, Hwanjin and Kim, Minje and Choi, Junil},
  booktitle={2025 IEEE Wireless Communications and Networking Conference (WCNC)},
  address = {Milan, Italy},
  year={2025},
}

@article{jiang2022accurate,
  title={Accurate channel prediction based on transformer: Making mobility negligible},
  author={Jiang, Hao and Cui, Mingyao and Ng, Derrick Wing Kwan and Dai, Linglong},
  journal={IEEE Journal on Selected Areas in Communications},
  volume={40},
  number={9},
  pages={2717--2732},
  year={2022},
  month={Sep.},
  publisher={IEEE}
}

@article{helmy2023lstm,
  title={{LSTM}-{GRU} model-based channel prediction for one-bit massive MIMO system},
  author={Helmy, Islam and Tarafder, Pulok and Choi, Wooyeol},
  journal={IEEE Transactions on Vehicular Technology},
  volume={72},
  number={8},
  pages={11053--11057},
  year={2023},
  month={Aug.},
  publisher={IEEE}
}

@INPROCEEDINGS{5947055,
  author={Palleit, Nico and Weber, Tobias},
  booktitle={2011 IEEE International Conference on Acoustics, Speech and Signal Processing (ICASSP)}, 
  title={Time prediction of non flat fading channels}, 
  address={Prague, Czech},
  year = {2011}
}

@INPROCEEDINGS{romano2019conformalized,
  author={Romano, Yaniv and Patterson, Evan and Candes, Emmanuel},
  booktitle={Advances in neural information processing systems}, 
  title={Conformalized quantile regression}, 
  address={Vancouver, Canada},  
year={2019}  
}

@inproceedings{min2007mimo,
  title={{MIMO}-{OFDM} downlink channel prediction for {IEEE}802. 16e systems using {K}alman filter},
  author={Min, Changkee and Chang, Namseok and Cha, Jongsub and Kang, Joonhyuk},
  booktitle={2007 IEEE Wireless Communications and Networking Conference},
  address={{H}ong {K}ong, China},
  year={2007}
}

@ARTICLE{10964108,
  author={Kim, Dongwon and Zecchin, Matteo and Park, Sangwoo and Kang, Joonhyuk and Simeone, Osvaldo},
  journal={IEEE Transactions on Signal Processing}, 
  title={Robust Bayesian Optimization via Localized Online Conformal Prediction}, 
  year={2025},
  volume={73},
  number={},
  pages={2039-2052}}

@inproceedings{kashyap2017performance,
  title={Performance analysis of {(TDD)} massive MIMO with Kalman channel prediction},
  author={Kashyap, Salil and Moll{\'e}n, Christopher and Bj{\"o}rnson, Emil and Larsson, Erik G},
  booktitle={2017 IEEE International Conference on Acoustics, Speech and Signal Processing (ICASSP)},
  address={{N}ew {O}rleans, USA},
  year={2017},
}

@ARTICLE{deng2025scan,
  author={Deng, Weicao and Shi, Binpu and Li, Min and Simeone, Osvaldo},
  journal={IEEE Transactions on Cognitive Communications and Networking}, 
  title={SCAN-BEST: Sub-6GHz-Aided Near-Field Beam Selection With Formal Reliability Guarantees}, 
  year={2026},
  month={Jan.},
  volume={12},
  number={},
  pages={5506-5521},
  doi={10.1109/TCCN.2026.3657091}}

@ARTICLE{6798744,
  author={Lu, Lu and Li, Geoffrey Ye and Swindlehurst, A. Lee and Ashikhmin, Alexei and Zhang, Rui},
  journal={IEEE Journal of Selected Topics in Signal Processing}, 
  title={An Overview of Massive MIMO: Benefits and Challenges}, 
  year={2014},
  month={Oct.},
  volume={8},
  number={5},
  pages={742-758},
  doi={10.1109/JSTSP.2014.2317671}}

@ARTICLE{10054381,
  author={Wang, Cheng-Xiang and You, Xiaohu and Gao, Xiqi and Zhu, Xiuming and Li, Zixin and Zhang, Chuan and Wang, Haiming and Huang, Yongming and Chen, Yunfei and Haas, Harald and Thompson, John S. and Larsson, Erik G. and Renzo, Marco Di and Tong, Wen and Zhu, Peiying and Shen, Xuemin and Poor, H. Vincent and Hanzo, Lajos},
  journal={IEEE Communications Surveys \& Tutorials}, 
  title={On the Road to 6{G}: Visions, Requirements, Key Technologies, and Testbeds}, 
  year={2023},
  month={2nd Quart.},
  volume={25},
  number={2},
  pages={905-974},
  doi={10.1109/COMST.2023.3249835}}

@ARTICLE{channel2015aging,
  author={Kong, Chuili and Zhong, Caijun and Papazafeiropoulos, Anastasios K. and Matthaiou, Michail and Zhang, Zhaoyang},
  journal={IEEE Transactions on Communications}, 
  title={Sum-Rate and Power Scaling of Massive MIMO Systems With Channel Aging}, 
  year={2015},
  month={Dec.},
  volume={63},
  number={12},
  pages={4879-4893},
  doi={10.1109/TCOMM.2015.2493998}}

@inproceedings{hoydis2023sionna,
  title={Sionna RT: Differentiable ray tracing for radio propagation modeling},
  author={Hoydis, Jakob and A{\"\i}t Aoudia, Fay{\c{c}}al and Cammerer, Sebastian and Nimier-David, Merlin and Binder, Nikolaus and Marcus, Guillermo and Keller, Alexander},
  booktitle={Proc. IEEE Globecom Workshops (GC Wkshps)},
  pages={317--321},
  year={2023},
  month={Dec.}
}

@inproceedings{kim2020massive,
  title={{M}assive {MIMO} channel prediction: {M}achine learning versus {K}alman filtering},
  author={Kim, Hwanjin and Kim, Sucheol and Lee, Hyeongtaek and Choi, Junil},
  booktitle={2020 IEEE Globecom Workshops (GC Wkshps)},
  address={Taipei, Taiwan},
  year={2020},
}

@techreport{specification3GPP,
  author       = "{3GPP}",
  title        = "{5G; Study on channel model for frequencies from 0.5 to 100GHz}",
  type         = "{TR}",
  number       = "38.901",
  institution  = "3GPP",
  version      = "v16.1.0",
  year         = "2020",
  month        = "11",
  note         = "Release 16",
}

@article{jiang2020deep,
  title={Deep learning for fading channel prediction},
  author={Jiang, Wei and Schotten, Hans Dieter},
  journal={IEEE Open Journal of the Communications Society},
  volume={1},
  pages={320--332},
  year={2020},
  publisher={IEEE}
}

@inproceedings{zecchin2025generalization,
  title={Generalization and informativeness of weighted conformal risk control under covariate shift},
  author={Zecchin, Matteo and Hellstr{\"o}m, Fredrik and Park, Sangwoo and Shitz, Shlomo Shamai and Simeone, Osvaldo},
  booktitle={2025 IEEE International Symposium on Information Theory (ISIT)},
  pages={1--6},
  year={2025},
  organization={IEEE}
}

@ARTICLE{8979256,
  author={Yuan, Jide and Ngo, Hien Quoc and Matthaiou, Michail},
  journal={IEEE Transactions on Wireless Communications}, 
  title={Machine Learning-Based Channel Prediction in Massive MIMO With Channel Aging}, 
  year={2020},
  month={May},
  volume={19},
  number={5},
  pages={2960-2973},
  doi={10.1109/TWC.2020.2969627}}
\end{document}